# Antiphase boundaries in Ni-Mn-Ga single crystal – experiment and model

O. Heczko, F. Maca, V. Drchal, L. Fekete, L. Straka, J. Zemen

*FZU – Institute of Physics of the Czech Academy of Sciences, Na Slovance 1999/2, 18200 Prague 8, Czech Republic*

**Abstract**

Thermally induced antiphase boundaries (APBs) in ferromagnetic, ordered Ni-Mn-Ga single crystal exhibit complex, irregular shapes and closed loops without any lattice plane preferences. The APBs were visualized on polished (100) surface using magnetic force microscopy (MFM) at the same location in parent cubic austenite and monoclinic martensite with uniaxial magnetic anisotropy. Based on ab-initio calculation we suggest that one APB curve with dark & light contrast consists of a pair of APB interfaces with narrow, only one-layer thick, core with structural partial B2' order in contrast to full $L2_1$ order of bulk. Calculated magnetic contrast using magnetostatic continuum simulations agrees well with MFM observation.

Antiphase boundaries (APBs) as planar structural defects naturally occur in ordered compounds which can affect the magnetic and structural properties [1-4]. To understand their origin and structure is thus vital for materials development. The single, planar APB can be created by plastic deformation [5] or can naturally arise upon the ordering process or transformation from disordered to ordered compound. Such thermally induced antiphase boundaries usually exhibit complex irregular shapes and finite thickness with disordered core [6,7]. In ferromagnetic compounds such regions can exhibit magnetic properties which differ from the surrounding bulk [8,9]. In Mn-containing Heusler alloys the core is expected to be antiferromagnetic due to increasing closeness of Mn atoms [10,11,12] in contrast with Fe-based alloys in which an increase of saturation magnetization was observed [6].

In Ni-Mn-Ga magnetic shape memory alloys the concentration of APBs can be influenced by heat treatment and increases with increasing rate of cooling [13]. It can provide new functionality by increasing magnetic coercivity producing semipermanent magnet which can be nullified by mechanical force [14, 15]. On the other hand, it causes unwanted slight deterioration of magnetically induced reorientation (MIR) due to modest increase of twinning stress [13].

The thermal APBs in Ni-Mn-Ga Heusler compound are formed upon the disorder-order transformation from B2' to $L2_1$ which occurs at about 1073K and it is weakly of the first order [16]. As only Mn and Ga atoms are involved in ordering, one can expect that the APB structure is formed by misplaced Mn and Ga atoms. The decreased distance between Mn atoms results in antiferromagnetic interaction between them [1, 11] and thus it can be reasonably expected that the magnetic ordering within APB is antiferromagnetic. This is supported by recent experiments using NMR [17] and by the study of the magnetization approach to saturation [10].

However, in ordered Heusler Ni-Mn-Ga alloys the precise structure of APBs and particularly the thickness of an APB region generating the characteristic dark & light APB contrast measured magnetic microscopy is hard to establish as there is no structural contrast in transmission electron microscopy (TEM) due to similarity of constituent atoms [18]. In contrast to the studies of ordered alloys containing rather different atomic species [19], the presence of APB is detected only by Lorenz TEM [20, 21] and magnetic force microscopy (MFM) microscopy [22, 23]. Since the APBs are revealed only by magnetic contrast (magnetic induction changes) the resolution is limited.

Therefore, in attempt to obtain a realistic atomistic model of APBs in Ni-Mn-Ga one should resort to modelling the boundary and compare it with the observation on a larger available scale. Here we present MFM observations of complex APBs in the same place of Ni-Mn-Ga single crystal in both phases, cubic austenite and modulated martensite. These observations agree well with the atomic simulation of APB and subsequent magnetostatic mesoscopic model.

Single crystal samples with composition Ni50Mn28Ga22 were cut in the shape of parallelepiped with {100} faces from single crystals. Surface was grinded, mechanically polished and electropolished by standard procedure [23]. The electropolishing to remove the surface layer affected by mechanical polishing is very important for visualization of antiphase boundaries (APB) [22]. The samples exhibited martensitic transformation to 10M modulated martensite at 310 K and magnetically induced reorientation (MIR) or magnetic shape memory effect at room temperature. A vibrating sample magnetometer (VSM) was used to measure magnetic properties and to confirm MIR in the sample. Applied field of 2 T was used to prepare the sample in single variant state with desired orientation of magnetization easy axis.

We used atomic force microscopy (AFM) to evaluate surface quality and then MFM to discern the APBs by magnetic contrast on the surface. Thus, we observe the APBs as surface projection on (100) plane. Imaging was performed at elevated temperature in cubic austenite and at ambient in modulated martensite. Moreover, due to MIR in 10M martensite we could observe the APBs in the same place for different orientations of easy magnetization axis.

Atomistic simulations of the APB region were carried out using Density Functional Theory (DFT). We assumed that the APB region consisted of a pair of planar antiphase interfaces surrounding the APB core. This configuration, where the APB region is formed by three Mn-Ga layers with magnetic moments opposite to the surrounding atoms, having the lowest ground state energy was used as an input of a magnetostatic calculation using the Finite Element Method (FEM) to predict magnetic contrast at length scales comparable to the MFM observation. The details of the calculations and further discussion of model justification are published elsewhere [25].

At room temperature, the single crystalline samples were in 10M modulated martensite state with magnetization easy axis along c-axis. As the uniaxial anisotropy of 10M martensite is relatively strong $1.7 \times 10^5$ J/m$^3$ [26] and with the c-axis in plane, the domain structure of the single variant, i.e., without twin boundary, is expected to exhibit parallel magnetic domains as shown in Fig. 1. The MFM figure shows several magnetic domains and strong single-color contrast (dark or light) of magnetic domain walls (DWs). The distinct contrast between domains with antiparallel magnetization and only approximate parallelism of DWs is due to slight deviation of surface from (100) plane caused by slight changes of crystal orientation upon the martensitic transformation from parent cubic to monoclinic lattice.

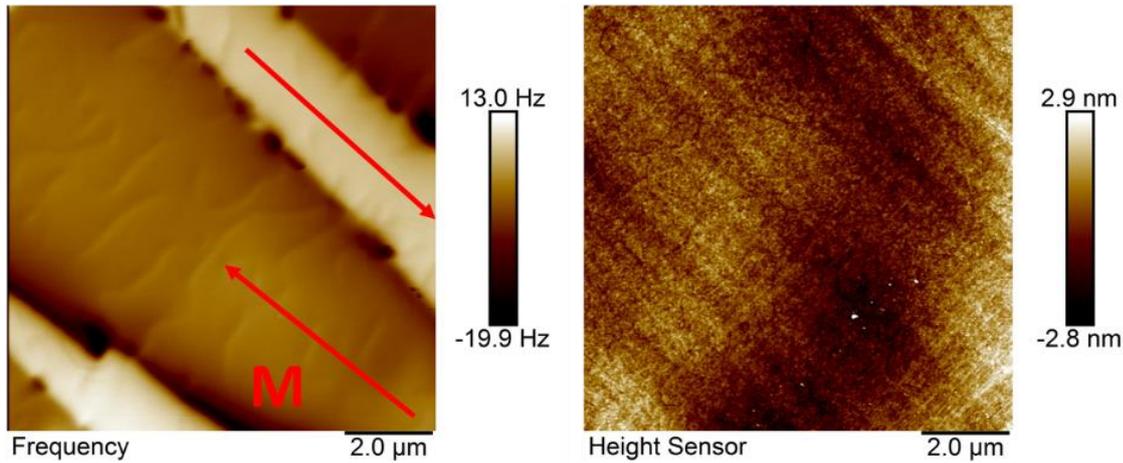

**Fig. 1** MFM and AFM figures of 10M martensite: a) magnetic domains and APBs in multidomain state, b) surface scan demonstrating the flatness of the surface.

Inside domains there are hardly discernible irregular curved lines which are ascribed to antiphase boundaries (APBs). These are better visible in the dark domains than in antiparallel (white) domains due to the high intensity of out-of-plane part of magnetization. Observed curvy lines are about perpendicular to magnetization in both domains in agreement with previous reports, where we demonstrated that only such contrast is observable [22]. Much stronger line contrast due to DW is irregular, which can be ascribed to the interaction with APBs. The presence of an APB changes the chirality of a DW and its contrast changes from dark to light and vice versa as the magnetization vector in the domain wall points out or into the surface [22, 27].

To avoid the inundation of weak contrast of APB in strong DW contrast, we prepared broad magnetic domains to observe APBs in single magnetic domain without DW interference. The much higher contrast of APBs in such single magnetic domain is shown in Fig. 2. Thanks to MIR the same sample can be prepared in two different structural orientations and while APBs stay in place, the magnetization direction changes to perpendicular one. This provides an additional line contrast which facilitates discerning APB shape or more precisely the line projection on (100) surface plane in full.

Another way to determine the full shape of APBs is to use the martensitic transformation. At elevated temperature the martensite transforms to cubic austenite. Although the APBs are invariantly observed in martensite due to strong magnetic anisotropy, the observation in austenite is much more difficult due to the very low magnetocrystalline anisotropy of austenite [28,29]. This causes very broad domains and the lack of well-defined magnetization direction which results in reduced contrast. On the other hand, the contours of APB are sometimes visible in full and not only in the direction perpendicular to the easy direction of magnetization. A reason can be inferred from ab-initio calculation, which suggests the preference of magnetization direction to be perpendicular to antiphase boundary [25]. The comparison of APBs in 10M martensite and austenite at the same place in the sample is shown in Fig. 2.

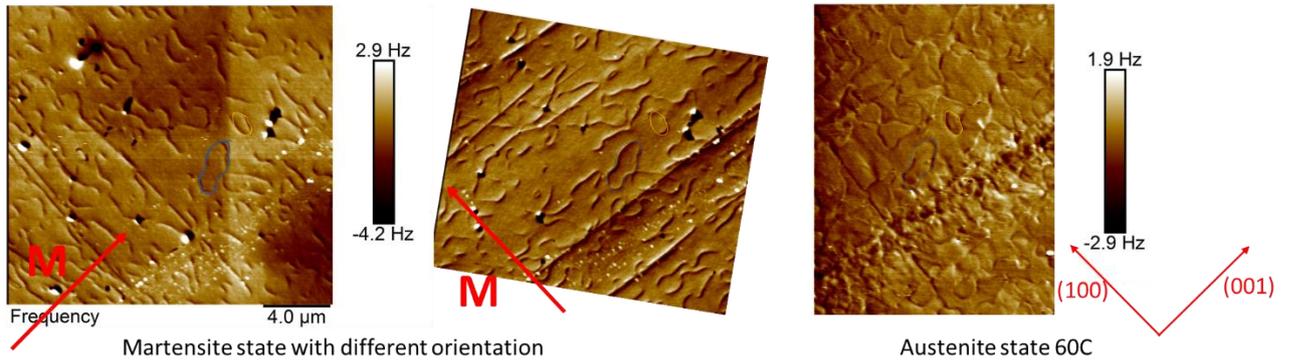

**Fig. 2.** MFM observation of APBs in: a) one twin martensite domain with two different orientations of easy magnetization axis (marked) and b) in parent phase austenite (cubic lattice and anisotropy is marked) obtained at elevated temperature. MFM micrographs are taken approximately in the same place. Two distinct APB domains are marked in all three figures (yellow and blue).

Significantly, the observed strong contrast on magnetic DW is a single line (unipolar) while double contrast (bipolar) is observed on APB boundary. This suggests that the thermally induced APB cannot be simple plane. If only one antiphase interface was present this would constitute atomically sharp magnetic domain wall in which magnetization abruptly changed to antiparallel direction [1, 30]. However, no such entity has been observed either in Lorentz TEM or MFM. It is confirmed by FEM simulations that such arrangement would result in a large magnetostatic energy penalty [25].

Based on the formation mechanism upon first order ordering transformation $L2_1$ – B2' [16] and available experimental data and literature hint [1] it seems that thermally induced APB region is made of two antiphase planes (interfaces), i.e., with Mn and Ga exchanged and region in-between (APB core) where Ga and Mn atoms are disordered or randomly mixed in their positions, That is the $L2_1$ order within antiphase domain changes to B2' partial order in the APB's core with Ni atom's position fixed. In addition, the observed APBs have complex shapes without apparent preference to any lattice plane reflecting the first order transformation proceeded by nucleation and growth [1,16].

However, the thickness of APB region (APB width) cannot be inferred from the observation. The MFM resolution is not sufficient to resolve any details of the APB. The LTEM suggests that the APB width is less than several nanometers [20,21] compared to the MFM contrast, which spreads over several tens of nm. Even TEM could not resolve further details. The direct observation of thermally induced APBs on different compounds in which the structure of APB can be resolved, shows narrow, irregularly curved regions with thin disordered core of few nm [7-9]. Similar APB width can be expected in our case.

Based on the observations and considering the way the thermally induced APBs are formed, we suggest a straightforward phenomenological model. The 1D schematic model along [001] direction within (100) plane of cubic L2$_1$ structure is shown in Fig. 3. The drawn thickness of the disordered APB core is just one plane thick as suggested by ab-initio calculation. Detailed discussion about ab initio calculation and possible other interpretations are published elsewhere [25]. The additional support for our model comes from recent NMR measurements [17] indicating that the APB boundary consists of a pair of antiphase interfaces surrounding one layer core.

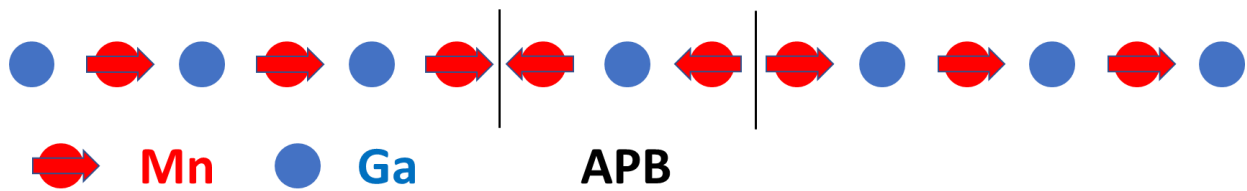

**Fig. 3.** Model of APB consisting of two sharp antiphase interfaces (marked by vertical lines) along [001] direction in (100) plane for Ni$_2$MnGa L2$_1$ cubic structure. Only Mn and Ga atoms are shown; Ni atoms are not visible as they are one layer out of plane.

Although the APB width seems to be too small to be observable, the arising profile of magnetic induction perpendicular to surface is apparently sufficient for MFM observation. The contrast modelled by magnetostatic continuum simulation is shown in Fig. 4 together with selected ultimate resolution MFM figure of several APBs. It shows that despite the one-layer thickness, the stray field perpendicular to the sample plane detected by MFM is strong enough for APB observation. The simulation also demonstrates that the contrast is visible only if the magnetization is about perpendicular to the APB. If the magnetization is parallel to APB line, no contrast arises in agreement with the experiment.

Moreover, the calculated contrast on APB is more than ten times weaker compared to stray field arising from single 180 deg magnetic domain. This agrees with the experimentally observed difference (Fig. 1). The full description of magnetostatic simulation and connection with atomistic models is published elsewhere [25].

The very small APB width can explain another puzzle, namely that magnetic coercivity is small and increases only modestly even for high density of APBs [13]. The coercivity in materials with high magnetocrystalline anisotropy is expected to be high due to the pinning on the obstacles. However, if the APBs regions are only ≈ 1 nm thick and the thickness of magnetic domain wall is more than 10 nm [21], the pinning is ineffective and thus the coercivity does not rise significantly in agreement with experiments.

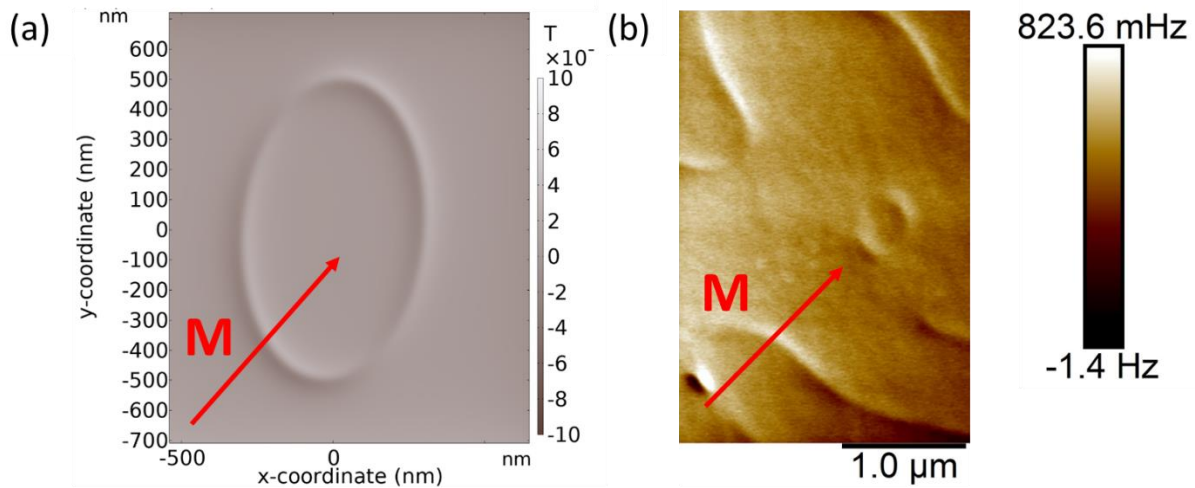

**Fig 4.** (a) APB contrast calculated using micromagnetic model (b) experimental MFM contrast on APB, please note small round feature. The direction of magnetization is marked.

Thermally induced APBs in ferromagnetic, ordered Ni-Mn-Ga compound were observed using magnetic force microscopy in parent cubic austenite at elevated temperature and in monoclinic martensite with uniaxial magnetic anisotropy at room temperature. The same place observations in austenite and martensite indicated complex, irregular and closed shape of APBs without any lattice plane preferences. The observation is in line with predictions of the multiscale model of APBs based on ab-initio atomistic and magnetostatic continuum simulations, which suggested that the APB region is composed of two antiphase interfaces surrounding the single-layer APB core.

**Acknowledgment**

This work was supported by the Czech Science Foundation (project No. 23-04806S) and by the Ferroic Multifunctionalities project (FerrMion) [Project No.

CZ.02.01.01/00/22008/0004591] by the Ministry of Education (MEY S), cofunded by the European Union. Computational resources were provided by the e−INFRACZ project (ID: 90254) supported by MEYS.

**Data availability statement**

The data that support the findings of this study are openly available in Zenodo at http://doi.org/[doi], reference number [reference number].

# References:

[1] A. J. Lapworth & J. P. Jakubovics, Effect of antiphase boundaries on the magnetic properties of Cu-Mn-Al Heusler alloys, Philosophical Magazine, 29, 253-273 (1974)

[2] Nedelkoski, Zlatko, Ana M. Sanchez, Arsham Ghasemi, Kohei Hamaya, Richard FL Evans, Gavin R. Bell, Atsufumi Hirohata, and Vlado K. Lazarov. "The antiphase boundary in half-metallic Heusler alloy Co2Fe (Al, Si): atomic structure, spin polarization reversal, and domain wall effects." *Applied Physics Letters* 109, no. 22 (2016).

[3] Ishikawa, H., R. Y. Umetsu, K. Kobayashi, A. Fujita, R. Kainuma, and K. Ishida. "Atomic ordering and magnetic properties in Ni2Mn (GaxAl1− x) Heusler alloys." *Acta Materialia* 56, no. 17 (2008): 4789-4797. https://doi.org/10.1016/j.actamat.2008.05.034

[4] Gomell, Leonie, Shyam Katnagallu, Abou Diack-Rasselio, Stefan Maier, Loic Perriere, Christina Scheu, Eric Alleno, and Baptiste Gault. "Chemical segregation and precipitation at anti-phase boundaries in thermoelectric Heusler-Fe2VAl." *Scripta Materialia* 186 (2020): 370-374.

[5] Ikeda, Kôki, and Seiki Takahashi. "Cold-working effect on magnetic properties in the Heusler alloys." *Physical Review B* 30, no. 7 (1984): 3808. DOI: https://doi.org/10.1103/PhysRevB.30.3808

[6] Murakami, Y., K. Niitsu, T. Tanigaki, R. Kainuma, H. S. Park, and D. Shindo. "Magnetization amplified by structural disorder within nanometre-scale interface region." *Nature communications* 5, no. 1 (2014): 4133.

[7] Niitsu, K., K. Minakuchi, X. Xu, M. Nagasako, I. Ohnuma, T. Tanigaki, Y. Murakami, D. Shindo, and R. Kainuma. "Atomic-resolution evaluation of microsegregation and degree of atomic order at antiphase boundaries in Ni50Mn20In30 Heusler alloy." *Acta Materialia* 122 (2017): 166-177.

[8] Yano, T., Y. Murakami, R. Kainuma, and D. Shindo. "Interaction between magnetic domain walls and antiphase boundaries in Ni2Mn (Al, Ga) studied by electron holography and Lorentz microscopy." *Materials transactions* 48, no. 10 (2007): 2636-2641.

[9] Murakami, Yasukazu, T. Yano, R. Y. Umetsu, R. Kainuma, and D. Shindo. "Suppression of ferromagnetism within antiphase boundaries in Ni50Mn25Al12. 5Ga12. 5 alloy." *Scripta Materialia* 65, no. 10 (2011): 895-898.

[10] Webster, P.J., 1969. Heusler alloys. *Contemporary Physics*, *10*(6), pp.559-577.

[11] Enkovaara, J., Heczko, O., Ayuela, A. and Nieminen, R.M., 2003. Coexistence of ferromagnetic and antiferromagnetic order in Mn-doped Ni 2 MnGa. *Physical Review B*, *67*(21), p.212405.

[12] Umetsu, R. Y., H. Ishikawa, K. Kobayashi, A. Fujita, K. Ishida, and R. Kainuma. "Effects of the antiferromagnetic anti-phase domain boundary on the magnetization processes in Ni2Mn (Ga0. 5Al0. 5) Heusler alloy." *Scripta Materialia* 65, no. 1 (2011): 41-44.

[13] Straka, L; Fekete, L; Rames, M; Belas, E; Heczko, O, Magnetic coercivity control by heat treatment in Heusler Ni-Mn-Ga(-B) single crystals, *Acta Mat*. 169 (2019) 109-121, 10.1016/j.actamat.2019.02.045.

[14] L. Straka, A. Soroka, O. Heczko, H. Hanninen, A. Sozinov, Mechanically induced demagnetization and remanent magnetization rotation in Ni-Mn-Ga(-B) magnetic shape memory alloy, Scripta Mater. 87 (2014) 25-28.

[15] Q. Peng, J. Huang, M. Chen, Q. Sun, Phase-field simulation of magnetic hysteresis and mechanically induced remanent magnetization rotation in Ni-Mn-Ga ferromagnetic shape memory alloy, Scripta Mater. 127 (2017) 49-53.

[16] Overholser, Ron W., Manfred Wuttig, and D. A. Neumann. "Chemical ordering in Ni-Mn-Ga Heusler alloys." *Scripta Materialia* 40 (1999): 1095-1102. https://doi.org/10.1016/S1359-6462(99)00080-9

[17] V. Chlan, M. Adamec, and O. Heczko. "Investigation of local surrounding of Mn atoms in Ni-Mn-Ga Heusler alloy using nuclear magnetic resonance." *arXiv preprint arXiv:2505.19968* (2025) to be submitted

[18] De Graef, Marc, Introduction to conventional transmission electron microscopy. Cambridge university press, 2003 and De Graef, Marc. "2. Lorentz microscopy: theoretical basis and image simulations." *Experimental methods in the physical sciences* 36 (2001): 27-67.

[19] Murakami, Y., K. Yanagisawa, K. Niitsu, H. S. Park, T. Matsuda, R. Kainuma, D. Shindo, and A. Tonomura. "Determination of magnetic flux density at the nanometer-scale antiphase boundary in Heusler alloy Ni50Mn25Al12. 5Ga12. 5." *Acta materialia* 61, no. 6 (2013): 2095-2101.

[20] Venkateswaran, S. P., N. T. Nuhfer, and M. De Graef. "Anti-phase boundaries and magnetic domain structures in Ni2MnGa-type Heusler alloys." Acta Materialia 55 (2007): 2621-2636.

[21] Vronka, Marek, Oleg Heczko, and Marc De Graef, Influence of antiphase and ferroelastic domain boundaries on ferromagnetic domain wall width in multiferroic Ni-Mn-Ga compound, *Applied Physics Letters* 115 (2019)

[22] L. Straka, L. Fekete, and O. Heczko, Antiphase boundaries in bulk Ni-Mn-Ga Heusler alloy observed by magnetic force microscopy, Applied Physics Letters 113 (2018).

[23] O. Heczko, Antiphase boundaries in Ni-Mn-Ga ordered compound, AIP Advances 10, 015137 (2020); https://doi.org/10.1063/1.5130183

[24] Heczko, Oleg, Aleksandr Soroka, and Simo-Pekka Hannula, Magnetic shape memory effect in thin foils, Applied Physics Letters 93 (2008).

[25] Jan Zemen, František Máca, Václav Drchal, Martin Veis, Oleg Heczko, Structure of Antiphase boundaries in Ni2MnGa: multiscale modeling, ArXiv 2025 to be published

[26] Straka, Ladislav, and Oleg Heczko. "Magnetic anisotropy in Ni–Mn–Ga martensites." Journal of Applied Physics 93, no. 10 (2003): 8636-8638.

[27] Vronka, M., Straka, L., De Graef, M. and Heczko, O., 2020. Antiphase boundaries, magnetic domains, and magnetic vortices in Ni–Mn–Ga single crystals. *Acta Materialia*, *184*, pp.179-186.

[28] Heczko, O., Kopeček, J., Majtás, D. and Landa, M., 2011, July. Magnetic and magnetoelastic properties of Ni-Mn-Ga–Do they need a revision?, Journal of Physics: Conference Series Vol. 303, p. 012081. IOP Publishing.

[29] Perevertov, A., I. Soldatov, R. Schaefer, R. H. Colman, and O. Heczko, "Kerr microscopy study of magnetic domains and their dynamics in bulk Ni-Mn-Ga austenite." *arXiv preprint arXiv:2506.23617* (2025) to be published in APL.

[30] Zweck, Ulrike, Pascal Neibecker, Sebastian Mühlbauer, Qiang Zhang, Pin-Yi Chiu, and Michael Leitner, Magnetization reversal induced by antiphase domain boundaries in Ni 2 Mn Z Heusler compounds, Physical Review B 106 (2022) 224106.